\documentclass[9pt,twocolumn,twoside]{osajnl}

\journal{ol} 

\setboolean{shortarticle}{true} 

\title{Pump-probe micro-spectroscopy by means of an ultra-fast acousto-optics delay line}

\author[1]{Xavier Audier}
\author[1]{Naveen Balla}
\author[,*]{Hervé Rigneault}

\affil[1]{Aix Marseille Université, CNRS, Centrale Marseille, Institut Fresnel UMR 7249, 13397, Marseille, France}

\affil[*]{Corresponding author: herve.rigneault@fresnel.fr}

\dates{accepted 8 December 2016}

\ociscodes{(300.6500) Spectroscopy, time-resolved; (230.1040) Acousto-optical devices; (320.5390) Picosecond phenomena; (180.4315) Nonlinear microscopy.}

\doi{\url{https://doi.org/10.1364/OL.42.000294}}

\begin{abstract}
We demonstrate femtosecond pump-probe transient absorption spectroscopy using a programmable dispersive filter as an ultra-fast delay line. Combined with fast synchronous detection this delay line allows for recording of 6~ps decay traces at 34~kHz. With such acquisition speed we perform single point pump-probe spectroscopy on bulk samples in 80~$\mu$s and hyperspectral pump-probe imaging over a field of view of 100$\mu$m in less than a second. The usability of the method is illustrated on a showcase experiment to image and discriminate two pigments in a mixture.
\end{abstract}

\setboolean{displaycopyright}{true}

\begin{document}

\maketitle

Fluorescence is a widely used contrast mechanism in spectroscopy~\cite{Lakowicz_83} and imaging~\cite{Pawley_06}, but it remains unable to address chromophores that exhibit poor or no fluorescence due to dominant non-radiative decay of their short-lived excited state~\cite{Turro_91}. Recent progress in multiphoton spectroscopy and microscopy has enabled the investigation of these non-fluorescent materials by means of ultra-fast pump-probe spectroscopy~\cite{Min_09} for uniquely identifying a variety of targets such as melanin~\cite{Ye_03,Fu_07} but also historical pigment~\cite{Samineni_12}. Implemented in a point-scanning microscopy scheme, pump-probe microscopy has been able to differentiate between eumelanin and pheomelanin~\cite{Matthews_11} with important application in melanoma diagnosis~\cite{Robles_15} but also to reveal the layered structure and color palette of historical paintings~\cite{Villafana_14,Villafana_16}.

In the most common implementation scheme~\cite{Turro_91,Min_09,Ye_03,Fu_07,Samineni_12,Matthews_11,Robles_15,Villafana_14,Villafana_16}, a first (pump) pulse interacts with the population of the ground and/or the excited states of the studied sample and a subsequent second (probe) pulse probes the transient interaction in the time domain. Varying the time delay $\delta\mathrm{t}$ between the pump and the probe pulses gives access to the excited state lifetime. The transient interaction for the probe pulse can be a gain of photons (an increase in transmission), a loss of photons (an increase in absorption), or a combination of the two, depending on the temporal evolution of the excited wave packet following excitation by the pump pulse. For instance, ground state depletion and stimulated emission lead to positive signal while excited state absorption and two-photon absorption result in a negative signal. Usually, regular time sampling measurements are performed where a mechanical translation stage that covers the full excited state relaxation time is used to generate $N$ regularly spaced delays. The total time to perform such a measurement scales with $N$, since it is usually limited by the time $\Delta\mathrm{t}$ (of the order of tens of ms) it takes to move the mechanical translation stage from one position to the next. In practice the mechanical translation stage is moved between images and it takes minutes to acquire 100$\mu$m x 100$\mu$m size pump-probe images with 1 pixel per micrometer and a 0.3~ps resolution over a time range of 10~ps, limiting applications where 3D investigation is necessary~\cite{Samineni_12,Matthews_11,Robles_15,Villafana_14}.

\begin{figure}[htbp]
\centering
\fbox{\includegraphics[width=7.5 cm]{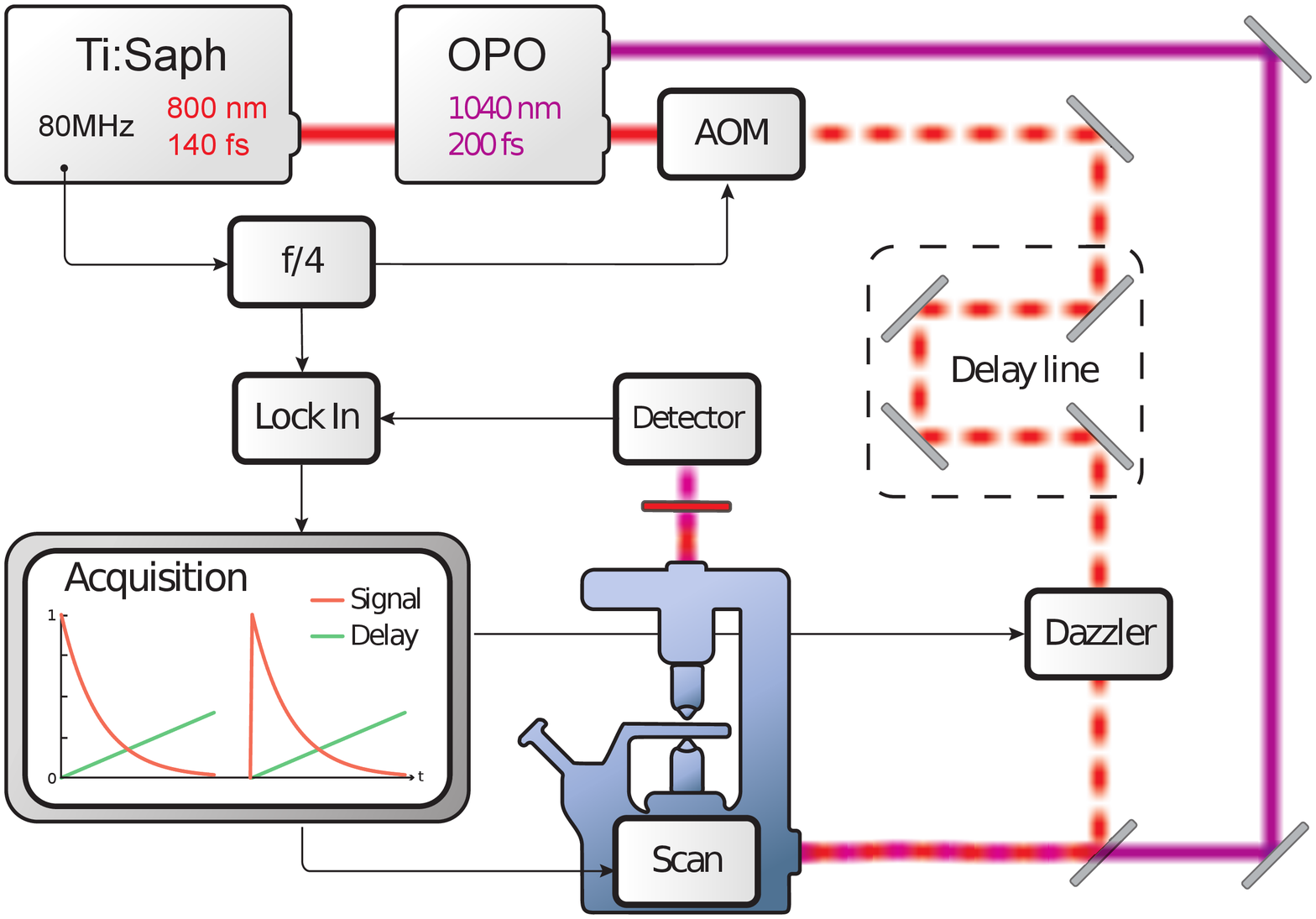}}
\caption{Experimental setup. The pump laser is modulated at one fourth (20~MHz) of the laser repetition rate. The acquisition software triggers an acoustic wave in the Dazzler, then records the extracted probe beam modulation as the delay between pump and probe pulses is rapidly swept.}
\label{fig:setup}
\end{figure}

A fast temporal scanning scheme is therefore direly needed to minimize acquisition speed, thermal drifts, and monitor rapidly evolving systems such as found in biology. Whereas temporal scanning can be achieved at few tens of Hertz using vibrating membranes over picoseconds time delays, it can be increased to few hundreds of Hertz using rotating mirrors~\cite{Domingue_14}. Dramatic temporal scanning speed improvement up to the kHz range have been reported using asynchronous optical sampling (ASOPS)~\cite{Elzinga_87} that employs two pulsed lasers with a small repetition rate offset to accomplish time-delay scanning without motion of any mechanical parts. Using ASOPS with 1~GHz repetition rate lasers, a record 5~kHz scan rate has been reported over 1~ns time delay with a temporal resolution limited by the laser pulse width (45~fs)~\cite{Gebs_10}. Even with high repetition rate lasers ASOPS is not well suited to probe ultra-fast molecular processes exhibiting a relaxation time of few ps; indeed in these cases ASOPS delay scanning goes up for few ns or more, as dictated by the laser repetition rate, and the useful data are collected during only about one-thousandth, or less, of the whole measurement time. Overcoming this limitation, all optical ultra-fast temporal scanning at 34~kHz over ps time delays has been recently reported using an acousto-optic programmable dispersive filter (AOPDF)~\cite{Schubert_13}. Recently the same group has illustrated the use of this technology in terahertz spectroscopy~\cite{Urbanek_16}. 

In this letter, we implement and demonstrate pump probe micro-spectroscopy and imaging using ultra-fast temporal scanning with an AOPDF. Transient absorption is recorded over a time delay of 6ps at the maximum frequency of 34~kHz. With this increased delay scanning speed, we were able to acquire temporally resolved pump-probe image stacks of a pigment mixture in less than a second; proving the relevance of this AOPDF pump-probe scheme for fast chemical species discrimination.


Figure~\ref{fig:setup} shows the schematic of the setup implemented to perform pump-probe micro-spectroscopy. The pump consists of a Ti:Sapphire laser (Chameleon{\texttrademark}, Coherent) providing 140~fs pulses at 800~nm with a repetition rate of 80~MHz. The probe is derived from an optical parametric oscillator (OPO) (Compact OPO-Vis, Coherent), pumped by the Ti:Sapphire laser, and providing 200~fs pulses at 1040~nm. Both beams are temporally synchronized as originating from the same pump laser. The pump beam is modulated in amplitude at one fourth of the laser repetition rate (20~MHz) via an acousto-optic modulator (MT200-A0.2-800, AA Opto Electronic). It is then sent through both the mechanical delay line and the AOPDF (Dazzler, Fastlite) before being recombined with the probe beam and send to a custom built inverted microscope equipped with galvo scanners (6215HM60, Cambridge Technology)~\cite{Brustlein_11,Ferrand_15}. Pump and probe beams are overlapped spatially and temporally at the focus of a NA=0.75 microscope objective lens (20x, Nikon) to illuminate and raster scan the sample. The probe beam is collect with a NA=1.15 microscope objective lens (40x, Nikon) filtered with a 850~nm long pass-filter (FEL0850,  Thorlabs) and send to a fast photodiode connected to a lock-in amplifier (LIA) (lock-in module, A.P.E). The LIA extracts the photodiode signal component at 20~MHz in phase with the modulation applied on the pump and its integration time is set to 300~ns to track rapid changes in modulation strength while fast scanning the delay with the AOPDF, as illustrated in figure~\ref{fig:setup}. In order to keep the pulse short at the sample plane the Dazzler dispersion is set to -25000~fs$^2$, a value that takes into account both the dispersion brought by the Dazzler acousto-optic material and the optics used in the setup.


Our choice of wavelengths, 800~nm and 1040~nm, is two fold. First, is allows to perform transient absorption measurement (pump-probe) on absorbing samples such as pigments. Second, it also targets CH$_2$ aliphatic chemical bounds at 2880~cm$^{-1}$ in stimulated Raman scattering (SRS), the pump and probe beams being the two SRS pump and Stoke pulses, respectively. SRS is an instantaneous process and is used here to measure the temporal overlap between pump and probe pulses.

\begin{figure}[htbp]
\centering
\fbox{\includegraphics[width=7.5 cm]{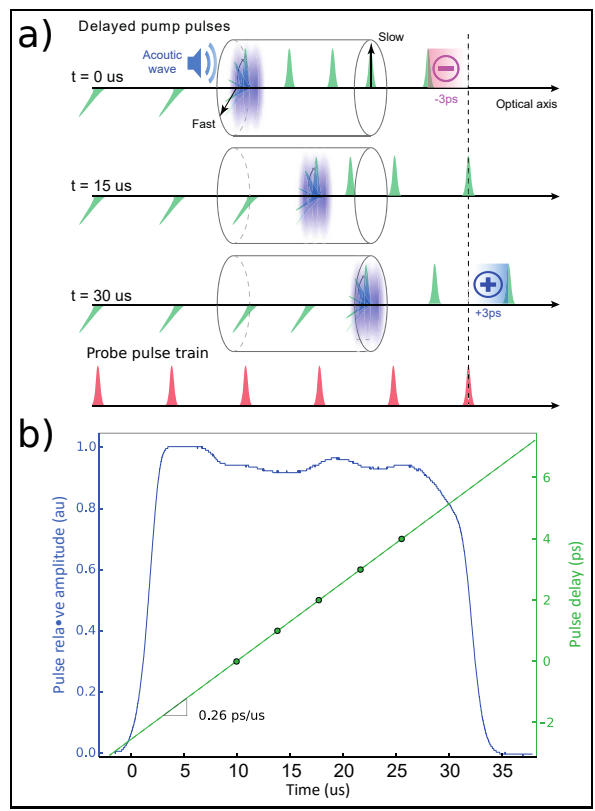}}
\caption{(a) Working principle of the Dazzler as a fast delay line. As the acoustic wave propagates inside a birefringeant crystal, it imprints a different delay on the laser pulses. (b)~Characterization of the pump pulses going through the Dazzler as the acoustic wave propagates inside. The solid blue line (left scale) is obtained by measuring the pump pulse directly. The green dots (right scale) are calibrated using a mechanical delay line. The dotted green line shows the linear fit of the data.}
\label{fig:dazzler}
\end{figure}

The Dazzler consists of a birefringent crystal in which an acoustic wave propagates collinear to the optical path~\cite{Verluise_00}. The acoustic wave rotates the laser polarization from the ordinary to the extraordinary axis. As illustrated in Figure~\ref{fig:dazzler}(a), when the acoustic wave propagates inside the crystal the incoming laser pulses will travel more distance subject to the ordinary (fast) index of refraction and less distance subject to the extraordinary (slow) index of refraction. This results in a continuous change of optical path that the successive laser pulses will be subject to, and therefore a change of delay between these pulses and a reference pulse train~\cite{Schubert_13}. The acoustic wave propagates inside the Dazzler crystal for 33~$\mu$s, delaying the pulses by a maximum of 6~ps.

\begin{figure}[tbp]
\centering
\fbox{\includegraphics[width=7.5 cm]{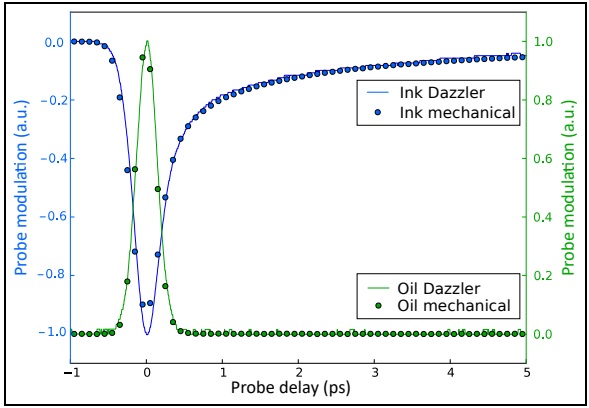}}
\caption{Nonlinear signals of oil (green) and black ink (blue) using the mechanical delay line (dots) or the Dazzler delay line (solid lines). The oil signal corresponds to the pulse temporal overlap as measured with stimulated Raman scattering. The black ink signal is characteristic of excited state absorption.}
\label{fig:liquids}
\end{figure}

To fully characterize the effect of the Dazzler on our pump beam we measure the intensity and the relative delay of the pulses at its output. To measure the intensity of the transmitted pulses we sent the modulated pump beam directly onto the detector. Figure~\ref{fig:dazzler}(b) (blue line) shows the demodulated signal from the LIA, assessing that the Dazzler has a roughly uniform transmission within the time it takes for the acoustic wave to propagate inside the crystal. In order to calibrate the amount of delay imprinted on the pulse by the Dazzler, we changed the delay of the pump using the mechanical delay line. The imprinted delay was measured by looking at the temporal shift of the measured pump-probe response. The signal we used for this calibration is the overlap of the pump and probe pulses, given by performing stimulated Raman scattering (SRS) in a bulk oil sample. This signal is exemplified in Figure~\ref{fig:liquids} (green). The perfect linear fit shown in Figure~\ref{fig:dazzler}(b) shows that the applied delay increases at a rate of 0.26~ps every microsecond that the acoustic wave propagates inside the crystal. This is consistent with the theory and already published measurements~\cite{Schubert_13}.


As a first step towards pump-probe imaging we realized pump-probe measurements in bulk samples. As explained previously, for our choice of wavelengths, SRS gives a positive (SR Gain) signal out of the LIA when the pump and probe (Stoke) pulses overlap in bulk olive oil (Figure~\ref{fig:liquids}, green).
For the second sample we used commercial black ink (Parker, Black ink cartridge) as an example of absorbing pigment. In this case, the pump laser excites the pigment molecules to higher energy states and the probe experiences excited state absorption. This result in a loss of probe signal, or opposite phase modulation, and negative signal after demodulation. The signal slowly goes to zero with increasing delay between pump and probe pulses, providing information on the way the excited molecules decay back to their ground states. This is the basis of time-resolved pump-probe spectroscopy.

In order to validate the usability of the Dazzler in this application, we compare the signals obtained using only a mechanical delay line and those obtained using the Dazzler as a fast delay line. Figure~\ref{fig:liquids} shows no difference between the two methods, therefore adding to the existing evidence that the AOPDF delay line can replace mechanical delay lines to increase acquisition speed. In particular, the rise-time and integration time of the LIA were sufficiently short to allow tracking of the signal modulation with no observable distortion compared to the mechanical delay approach.

The main advantage of the AOPDF approach is its acquisition time. For the mechanical delay line most of the time is spent moving the stage which takes seconds. For the Dazzler a full measurement only takes 33~$\mu$s, a five orders of magnitude improvement in speed as compared to the mechanical delay stage.
On the other hand, the range of delay accessible (6~ps) and the speed at which the delay is swept (33~$\mu$s) are set by the properties of the AOPDF crystal and cannot easily be tuned. The AOPDF system is also bounded to sweep delays and cannot stop at a particular one. Finally the diffraction efficiency of the Dazzler is acceptable ($\sim$~30~\%) but one should also anticipate power loss due to the presence of the Dazzler delay line.


We further use the Dazzler delay line to acquire pump-probe images of pigments. For each pixel a single acoustic wave is sent in the Dazzler crystal, and one pump probe response is acquired during 80~$\mu$s. A data acquisition card collects the output of the LIA every 800~ns. The focal spot is scanned over a 100~$\mu$m square field of view with a 1~$\mu$m pixel size. The final image is a 100 x 100 pixels by 100 delays stack acquired in 0.8 second. More precise synchronization between the acoustic wave and the acquisition card could ultimately reduce this time to 0.33 second.
\begin{figure}[htbp]
\centering
\fbox{\includegraphics[width=7.5 cm]{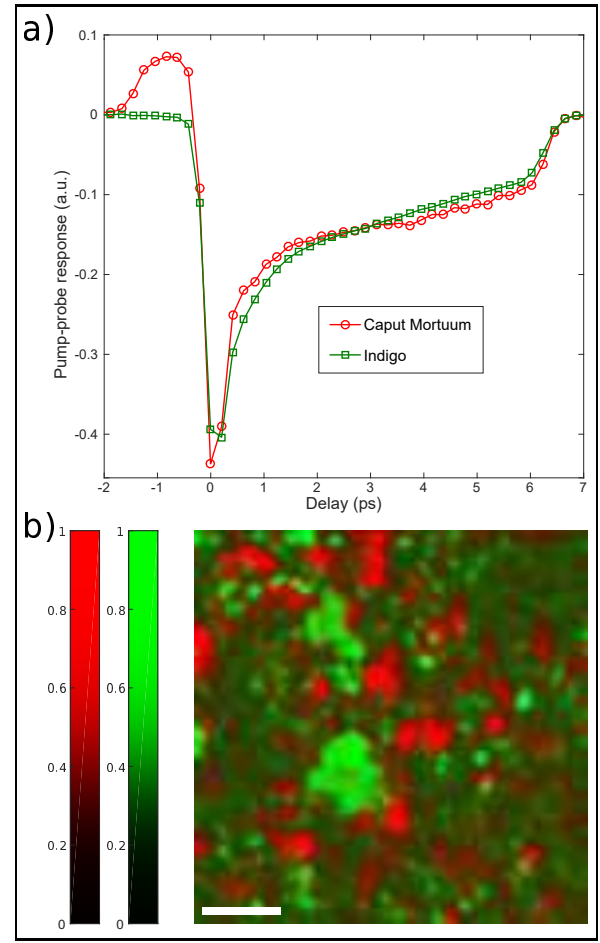}}
\caption{(a) Pump-probe image of a mixture of two pigments. (b)~Hyperspectral image stack projected on components of Caput Mortuum (red) and Indigo (green). The scale bar is 20~$\mu$m. The concentrations are normalized.}
\label{fig:imaging}
\end{figure} 
We apply this fast pump-probe scheme to image pigments found in paintings~\cite{Samineni_12}. Here we used Caput Mortuum (Kremer pigments 48750) and Indigo (Kremer pigments 36000). These two pigments exhibit relatively close pump-probe signals for the selected wavelengths making their separation challenging. The pump-probe signal characteristic of each pigments is shown in Figure~\ref{fig:imaging}(a). It was extracted by averaging the signal from a 100~$\mu$m square field of view containing the pigment alone.
Caput Mortuum shows an unusual positive signal for delays below zero. This is likely due to some decay paths having lifetimes of nanoseconds. In this case the modulation transfer happens between the probe and the previous pump pulse that arrived on the sample 12.5~ns before. Although this does not cause any problem with our imaging scheme, we restricted the delay times to only positive values during the later image analysis to keep the main focus on picosecond lifetimes.

We now use our Dazzler based fast pump-probe imaging scheme to identify the two above mentioned pigments (Caput Mortuum and Indigo) in a sample containing a mixture of the two. The image was acquired in the exact same conditions as previously described. The only image processing applied was a Gaussian filter with one pixel variance to spatially smooth the images. In the image stack, each pixel is a vector consisting of a pump-probe measurement for different delays. In order to extract the concentration of the species present in this pixel we use unconstrained demixing~\cite{Chang_07}. This method projects the recorded vector on the vectors corresponding to the known species. The output of this demixing is a concentration map giving the quantity of each species in the field of view. Figure~\ref{fig:imaging}(b) shows the concentration maps associated with Caput Mortuum (red) and Indigo (green) in the pigment mixture. We identify non overlapping domains corresponding to a clear separation of the pigment grains or aggregates. Here again the full hyperspectral image (100~pixels x 100~pixels x 100~delays) is acquired in 0.8~s, an order on magnitude faster than when using a mechanical delay stage. The Dazzler based pump-probe imaging also inverts the temporal and spatial acquisition timescales as compared to the mechanical delay based approach: for each pixel we scan the different delays before moving to the neighboring pixel to reconstruct the image. The advantage of this technique is that there is minimal dead times in the acquisition and no problem of pixel shift for different delays, a recurrent problem that seriously hampers pump-probe imaging on moving samples found in life sciences.

Here we used a 200~fs probe pulse, this requires approximately a 2~$\mu$s acoustic wave duration into the Dazzler (including the dispersion compensation for the setup) and leaves free 29~$\mu$s (that is 7.5ps) for the delay scan. The Dazzler would allow to use shorter pulses to probe molecular relaxation dynamic with a better resolution without scarifying too much the delay scan range; for instance a 100~fs pulse would require a 4~$\mu$s acoustic wave leaving free 25~$\mu$s (6.5~ps) to perform the delay scan. Ultimately a 30~fs pulse would leave free 10~$\mu$s to perform a 2.6~ps delay scan. Nevertheless short pulses (30~fs) would require further care as the Dazzler induces a pulse duration that varies with the delay. This is because of the difference in dispersion between the ordinary and the extraordinary axis in the acousto-optic material, an effect that would alter the pump-probe time resolution across the relaxation decay. Schubert and coworkers~\cite{Schubert_13} have reported that this effect is not significant for pulses as short as 100~fs; for shorter pulses a more detailed analysis should be performed.


To conclude, we have demonstrated the use of an AOPDF to enhance the acquisition speed in pump-probe spectroscopy and imaging. The speed enhancement is about five orders of magnitude for single point measurements and more than one order of magnitude for imaging. This speed improvement opens several interesting perspectives. Single point spectroscopy on transient or evolving molecular processes could be performed with a resolution of few tens of $\mu$s, whereas pump-probe imaging becomes possible in living systems as the full decay time trace is acquired during one pixel dwell time. Other light matter interaction processes can benefit from a fast delay line, this is for instance the case of impulsive vibrational spectroscopy~\cite{Ogilvie_06} or spectral focusing coherent Raman scattering microscopy~\cite{Hellerer_04,Andresen_11}.

\section*{Funding Information}
Agence National de la Recherche (ANR11-INSB-0006, ANR-10-INSB-04-01 ); Fondation Universitaire A*MIDEX (ANR-11-IDEX-0001-02)

\section*{Competing financial interests}
The authors declare no competing financial interests.
 
\section*{Acknowledgments}
X. A. acknowledges the \'{E}cole Normale Supérieure for supporting his funding

\bigskip



\end{document}